# Acoustic Emission Sensor Network Optimization Based on Grid Loop Search and Particle Swarm Source Location

Yiling Chen[1], Xueyi Shang[2,*], Yi Ren[3], Linghao Liu[1], Xiaoying Li[4], Yu Zhang[5], Xiao Wu[6], Zhuqing Li[4] and Yang Tai[1]

1 School of Resources and Safety Engineering, Chongqing University, Chongqing 400044, China

2 State Key Laboratory of Strata Intelligent Control and Green Mining Co-founded by Shandong Province and the Ministry of Science and Technology, Shandong University of Science and Technology, Qingdao 266590, China

3 China Academy of Safety Science and Technology, Beijing 100012, China

4 Beijing 5lWORLD Digital Twin Technology Co., Ltd. Beijing 100080, China

5 Y-Link (Wuhan) Technology Co., Ltd. Wuhan 430010, China

6 Chongqing Top-Tech Information Co.,Ltd. Chongqing 400014, China

**Abstract:** The layout of acoustic emission sensors plays a critical role in non-destructive structural testing. This study proposes a grid-based optimization method focused on multi-source location results, in contrast to traditional sensor layout optimization methods that construct a correlation matrix based on sensor layout and one source location. Based on the seismic source travel-time theory, the proposed method establishes a location objective function based on minimum travel-time differences, which is solved through the particle swarm optimization (PSO) algorithm. Furthermore, based on location accuracy across various configurations, the method systematically evaluates potential optimal sensor locations through grid search. Synthetic tests and laboratory pencil-lead break (PLB) experiments are conducted to compare the effectiveness of PSO, genetic algorithm, and simulated annealing, with the following conclusions: (1) In synthetic tests, the proposed method achieved an average location error of 1.78 mm, outperforming that based on the traditional layout, genetic algorithm (GA), and simulated annealing (SA). (2) For different noise cases, the location accuracy separately improved by 24.89% ($\sigma=0.5\mu s$), 12.59% ($\sigma=2\mu s$), and 15.06% ($\sigma=5\mu s$) compared with the traditional layout. (3) For the PLB experiments, the optimized layout achieved an average location error of 9.37 mm, which improved the location accuracy by 59.15% compared with the Traditional layout.

**Keywords:** Acoustic Emission; Sensor Layout; Source Location; Particle Swarm Optimization

# 1 Introduction

Acoustic emission (AE) is a phenomenon where strain energy is released as transient elastic waves, which occurs when a material experiences localized plastic deformation or crack formation [1]. AE monitoring technology is an effective non-destructive monitoring method. It captures elastic waves through sensors [2] and allows for the determination of AE source location, source mechanism, and structural velocity. The sensor layout in an AE monitoring system is crucial for ensuring parameter high-precision inversion [3]. Therefore, it is necessary to develop an optimization method for the sensor network layout.

Early sensor layouts were determined based on personal experience. As a result, the sensor locations were highly subjective and often led to inaccurate results. In 1977, Kijko [4,5] proposed a D-value model for microseismic sensor layout optimization. This model determines the optimal sensor configuration by minimizing the determinant of the covariance matrix of source location and travel time. Due to its flexibility, this method has been widely applied in the design of sensor network layouts. In 1990, Rabinowitz and Steinberg [6] introduced the DETMAX method, which improved the D-value model by preventing two sensors at the same location. Although the D-value model provides an evaluation criterion for sensor layout design, it has some limitations: When the number of sensors is large, the model cannot cover all sensors and involves complex calculations. With the development of optimization algorithms, researchers have attempted to integrate these algorithms into sensor layout optimization [7]. For example, by employing genetic algorithms (GA) for sensor layout optimization, Jones [8] proposed an optimal three-dimensional microseismic monitoring network based on arbitrarily shaped cavities. In 2013, Kraft et al. [9] proposed a regional-scale microseismic monitoring network optimization method based on simulated annealing (SA) and the D-value optimization criterion. This method can determine the geometry of a monitoring network that meets specified requirements.

In summary, current sensor layout optimization primarily involves constructing an information matrix based on sensor locations and source location results. Optimization algorithms are then applied to the covariance matrix to obtain the optimized sensor layout. However, these approaches entail significant complexity and limited flexibility for adjustment. Therefore, the optimized sensor layout computation can be shifted to directly reduce source location errors, which can simplify the overall process and improve computational efficiency.

The focus is now placed on reducing AE location errors. To achieve this, the key lies in employing more advanced methods to solve the objective function for source location. The commonly used methods currently include gradient descent approaches, such as Newton's iteration and the simplex method [10,11] , as well as heuristic methods, including GA [12], particle swarm optimization (PSO) [13], and SA [14]. Gradient descent approaches are highly dependent on local derivative information and require highly accurate traditional values. In contrast, heuristic methods do not rely on derivative information and are less likely to become trapped in local optima. As a result, these methods are widely applied to solve the objective function in source location problems [15]. Among various heuristic algorithms, the PSO algorithm has been widely applied in geophysical inversion [16]. This algorithm is based on swarm intelligence and exhibits strong global optimization capabilities. For example, Chen et al. [17] proposed a microseismic source location method based on PSO and a layered velocity model, which achieved high location accuracy.

In summary, existing sensor layout optimization methods often apply optimization algorithms directly to sensor layout optimization, determining sensor locations by maximizing the determinant of the information matrix. This process is relatively complex. To address this, this study proposes an optimization method for sensor layout based on the minimization of travel-time differences. This study's main innovations and contributions are: (1) Proposing a minimum travel-time-difference based objective function for source location. It integrates PSO and grid search to optimize sensor layout, which improves the efficiency of location and enhances the method's interpretability. (2) Synthetic tests compared the proposed method with GA and SA. The proposed method improves location accuracy by 7.81% compared to GA and by 7.39% compared to SA algorithms. (3) For different noise conditions, the proposed method improved location accuracy by 24.89% ($\sigma=0.5$ μs), 12.59% ($\sigma=2$ μs), and 15.06% ($\sigma=5$ μs) compared to the traditional layout. These results demonstrate that the proposed method can maintain effective layout optimization under noise interference. (4) In laboratory AE pencil-lead break (PLB) experiments, the proposed method improved location accuracy by 59.15% compared to the traditional layout. It also achieved improvements of 32.18% and 20.48% compared to the genetic algorithm and the SA algorithm.

The remaining sections of this study are organized as follows: Section 2 introduces the construction and solution methods of the AE source location objective function. It also establishes a

PSO and grid search method combined sensor network layout optimization method. Section 3 generates synthetic data based on a planar rectangular sample. It evaluates the layout optimization performance of preset sensor configurations using different optimization algorithms. Section 4 conducts laboratory AE PLB experiments to test the proposed method. Section 5 provides a summary and outlook of this study.

## 2 Methodology

### 2.1 Source location objective function

Assume that the sample where AE events occur is isotropic. In this case, signals from AE sources propagate linearly through the medium at a constant velocity. Figure 1 shows the locations of AE events and sensors.

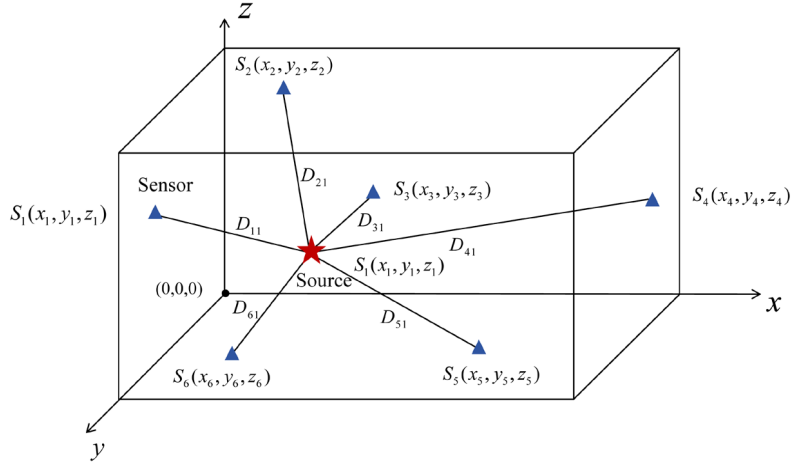

**Figure 1.** Schematic diagram of AE event and sensor locations

Assume that the P-wave velocity is $v_p$. Then, the distance $D_{ij}^{est}$ between the AE source $S_j(x_j, y_j, z_j), j = 1, 2, 3, ..., N$ and the sensor $S_i(x_i, y_i, z_i), i = 1, 2, 3, ..., M$ can be expressed as:

$$D_{ij}^{est} = \sqrt{(x_i - x_j)^2 + (y_i - y_j)^2 + (z_i - z_j)^2} \tag{1}$$

where $i$ represents the sensor id, and $j$ denotes the AE event id. Furthermore, based on Eq.(1), the travel time fitting error between the sensor and the AE source can be expressed as:

$$f_{ij} = \left| \frac{D_{ij}^{est}}{v_p} - t^{obser} \right| \tag{2}$$

where $t^{obser}$ represents the actual travel time for the distance between sensor $i$ and AE source $j$.

From Eq.(2), it can be observed that the closer $f_{ij}$ is to zero, the more accurate the location result of the AE source. Therefore, the location problem can be transformed into an optimization problem:

$$F(X) = \min \sum_{i=1}^{N} \sum_{j=1}^{M} f_{ij} \tag{3}$$

Therefore, the optimal AE source location can be determined by solving Eq.(3). The following section will introduce how to solve Eq.(3) by using the PSO algorithm.

**2.2 PSO based location algorithm**

PSO can efficiently search for the global optimum, which is suitable for solving complex problems. With the simple modeling and ease of implementation, PSO has been widely used in AE event locations. In this study, we employed PSO to solve the location objective function based on minimizing travel time difference Eq(3). The algorithm finds the optimal solution by simulating collaboration in a flock, initializing random solutions, and iteratively updating each particle's location and velocity. Each particle's movement is determined by its personal best location and the swarm's global best location, then the trajectory is adjusted to locate the optimal solution. The specific steps of PSO algorithm are as follows:

(1) Parameter initialization: Multi-particles are randomly generated, each particle represents a potential AE event location. Each particle is assigned a random velocity $V_k(0)$ and location $X_k(0)$, where $k$ represents the index of different particles.

Then, these initial parameters of each particle are substituted into the objective function Eq.(3) to calculate the initial fitness $F(X_k(0))$, which is then set as each particle's personal best $pbest_k(0)$.

(2) Velocity and location update: The velocity of each particle is updated according to Eq.(5):

$$V_k^{(t+1)} = w \cdot V_k^{(t)} + c_1 \cdot r_1 \cdot (pbest_k(t) - X_k^{(t)}) + c_2 \cdot r_2 \cdot (gbest_k(t) - X_k^{(t)}) \tag{4}$$

where $V_k^{(t+1)}$ represents the velocity of particle $k$ at time $t+1$, and $i$ is set to 30; $w$ is the inertia weight, which determines the particle's tendency to maintain its current velocity; $V_k^{(t)}$ is the velocity of particle $k$ at time $t$; $c_1$ and $c_2$ separately are learning factors that control the particle's attraction to its personal best and the global best location, and they are both set to 1.49; $r_1$ and $r_2$ are random numbers within the range [0,1].

Then, the particle's location is updated based on the updated velocity:

$$X_k^{(t+1)} = X_k^{(t)} + V_k^{(t+1)} \tag{5}$$

where $X_k^{(t+1)}$ represents the location of particle $k$ at time $t+1$, $X_k^{(t)}$ is the location of particle $k$ at time $t$.

(3) Fitness evaluation: After updating each particle's location $X_k$, the parameters represented by this location are substituted into the objective function $F$. The $F(X_k^{t+1})$ is used to update the individual historical best location $pbest_k$ of particle $k$ at time $t+1$ and the global best location $gbest(t+1)$ of the swarm.

(4) Updating $pbest_i$ and $gbest(t+1)$: The fitness function is used as a criterion, where lower values indicate higher accuracy in determining the optimal location. For each particle $i$, if the new fitness $F(X_k^{(t+1)})$ is smaller than the fitness $F(pbest_k(t))$ of its historical best location, the historical best location is updated as $pbest_k(t+1) = X_k^{(t+1)}$. Otherwise, the historical best location remains unchanged. Among the historical best locations $pbest_k(t+1)$ of all particles, the location with the best fitness is selected as the global best location $gbest(t+1)$ of the swarm. The updated equations are as follows:

$$pbest_k(t+1) = \begin{cases} X_k^{(t+1)}, & F(X_k^{(t+1)}) < F(pbest_k(t)) \\ pbest_k(t), & else \end{cases} \tag{6}$$

$$gbest(t+1) = \begin{cases} pbest_k(t+1), & F(pbest_k(t+1)) < F(gbest(t)) \\ gbest(t), & else \end{cases} \tag{7}$$

(5) Iteration and convergence: By iteratively updating each particle's velocity and location, updated $pbest_k$ and $gbest(t+1)$ are generated until the maximum number of iterations is reached. Ultimately, the global best location $gbest(t+1)$ is considered as the optimal coordinate for the AE event. This location represents the solution that minimizes the error of the objective function.

**2.3 Sensor layout optimization method**

PSO algorithm is used to evaluate the quality of the current sensor layout based on location errors of different sensor layouts. Sensor locations are adjusted step by step to reduce location errors. This study applies a grid search method to iteratively change sensor locations and calculate location errors for different configurations to optimize the sensor layout. Figure 2 presents the overall process of sensor layout optimization.

(1) Grid Partitioning: Within the sensor deployment area, a regular grid is generated based on predefined spacing. The density of grid points is determined by the required optimization accuracy. Assuming the target area is $L \times M$ mm², then the set of grid points can be expressed as:

$$\Omega = \{(x, y) \mid x \in \{x_1, x_2, ..., x_m\}, y \in \{y_1, y_2, ..., y_n\}, z \in \{z_1, z_2, ..., z_k\}\} \tag{8}$$

where $x_m$, $y_n$, and $z_k$ represent the coordinates of the grid points.

(2) Objective function evaluation: For each sensor $S_i(x_i, y_i, z_i), i = 1, 2, 3, ..., M$, the location is moved to each candidate location $\theta_k \in \Omega$. For each candidate location, the AE source location $P'$ is calculated using Eq.(3).

(3) Location error calculation: For each candidate location $\theta_k$, the location error $E$ is calculated as the distance between the AE source location $P'$ and the actual source location $P$:

$$E = \sqrt{(x_j - x)^2 + (y_j - y)^2 + (z_j - z)^2} \tag{9}$$

where $(x_j, y_j, z_j)$ is the source location, and $(x, y, z)$ is the actual source location.

(4) Update sensor locations: In each iteration, the candidate location $S_i$ with the minimum location error for each sensor $\theta_k^*$ is selected. This location is then updated as the new sensor layout configuration:

$$\theta_k^* = \min_{\theta_k \in \Omega} E \tag{10}$$

If the error at this location is smaller than the current global minimum error $E_{global}$, the global optimal location and the corresponding error are updated.

(6) Completion of sensor layout optimization: During the optimization process, the optimal sensor locations and the corresponding location errors for each iteration are recorded. The final optimization results are validated to ensure the rationality of the sensor layout and the improvement in location accuracy.

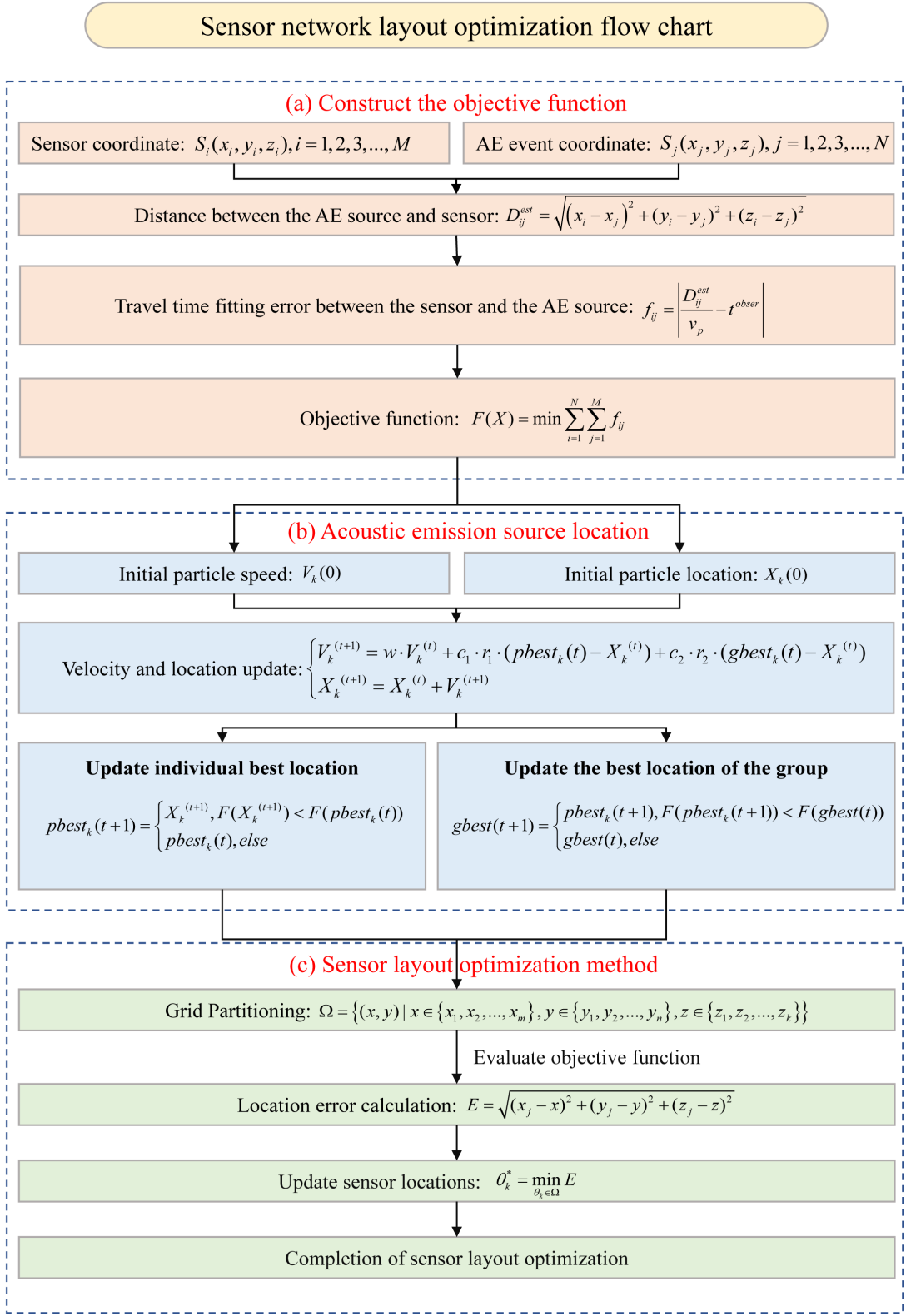

**Figure 2.** Grid search optimization method for sensor network layout

# 3 Synthetic test

## 3.1 Sensor network optimization

A sample with dimensions of 300 mm×200 mm was prepared. The P-wave velocity was set to $v_p$ =6 km/s, and the standard deviation of the arrival time error was assumed to be σ =0.5μs. Based on experimental experience, the sample was divided into a grid with a unit size of 10 mm × 10 mm. The grid intersection points were considered as potential sensor locations. To facilitate the implementation of subsequent indoor PLB experiments, we reserved a 25 mm margin around the rock sample. The potential sensor locations and the distribution of AE sources are shown in Figure 3.

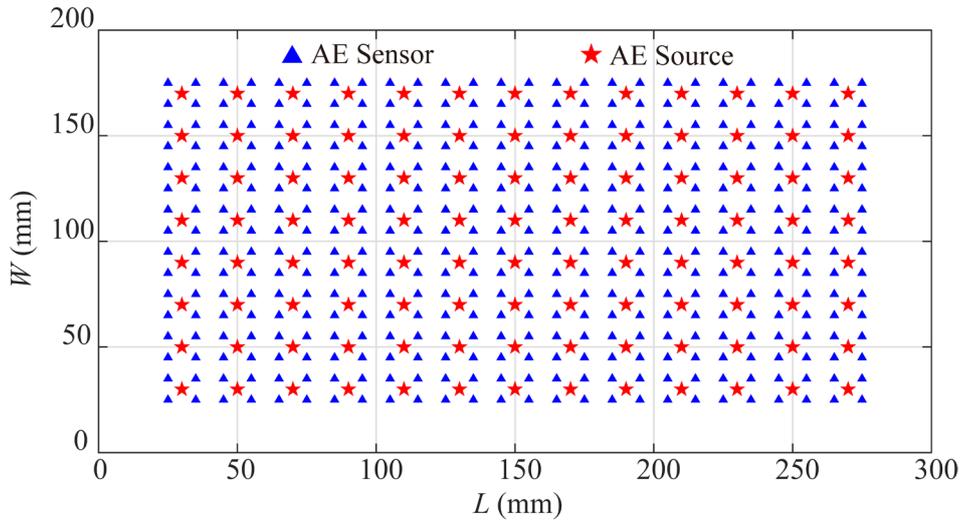

**Figure 3.** Potential AE and sensor locations in the monitoring system

Based on this setup, synthetic tests on sensor layout optimization were conducted. Eight sensors were used in this test. Traditional sensor layouts are often regular and symmetric. Therefore, to simulate traditional layouts, the traditional layout was designed with regular symmetry. In order to conduct comparative experiments with the PSO, we also recorded the sensor layout optimization results obtained using the Genetic Algorithm (GA) and SA (SA) for source localization solutions. The methods combining PSO, GA, and SA with the grid search method were defined as the PSO-GRID method, GA-GRID method, and SA-GRID method, respectively. The sensor distributions obtained by different methods under the same initial layout are shown in Figure 4. Compared to the traditional layout, the sensor distributions optimized by these three methods tend to concentrate along the boundaries of the monitoring area, while maintaining some sensors in the central region.

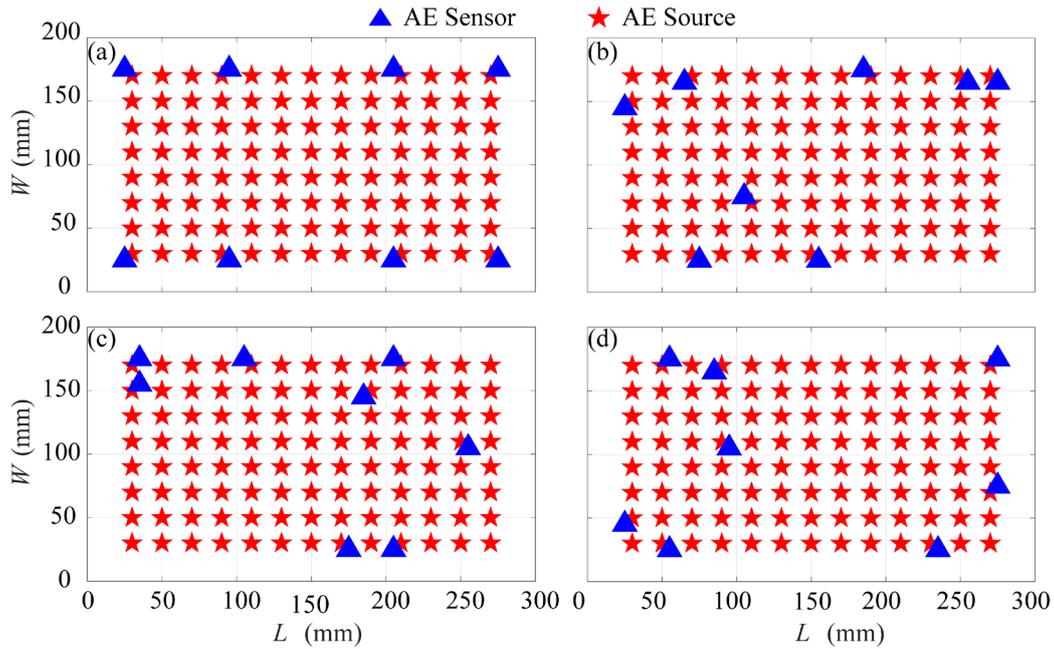

**Figure 4.** Sensor layouts before and after optimization. (a) Traditional sensor layout. (b-d) Sensor layout after optimization using the PSO-GRID method, GA-GRID method, and SA-GRID method.

During the optimization process, the global error sums of these three methods (Figure 5) were recorded to evaluate their performance in sensor layout optimization. The GA-GRID method demonstrates rapid convergence in the early stage, achieving a lower error value more quickly. Although the PSO-GRID method has a slightly slower convergence than the GA-GRID method, it achieves a slightly lower final error value than the GA-GRID method. In contrast, the SA-GRID method has a slower convergence rate and a higher final location error. The result indicates that the performance of the SA-GRID method is not as effective as the other two algorithms.

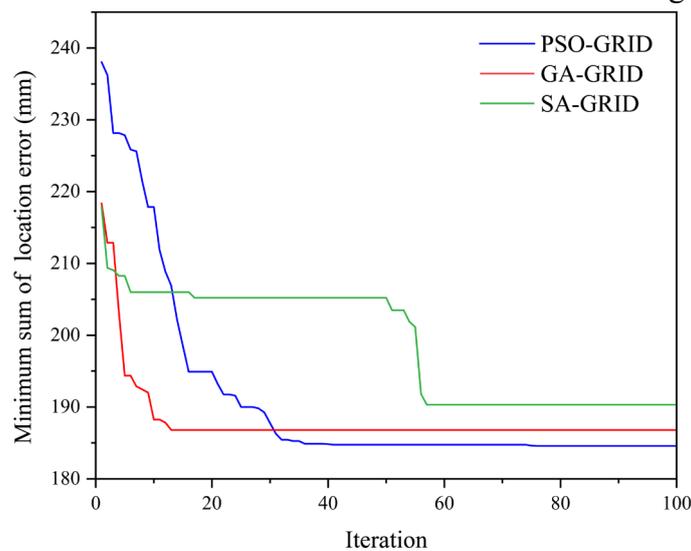

**Figure 5.** Global minimum error convergence for the PSO-GRID method, GA-GRID method, and SA-GRID method.

**3.2 Location results**

To analyze the performance of different optimization algorithms during the sensor layout optimization process, we recorded the location error values $E_i$ of all sensors under different layouts at each iteration (Figure 6). The diagonal elements of the matrix scatter plots show the distribution histograms of individual sensors during the optimization process, while the off-diagonal elements display the scatter plots of error values between different sensors.

By comparing these three methods, it is evident that the error distribution in the matrix scatter plot of the PSO-GRID method is more dispersed than that of GA and SA. This dispersed distribution indicates that PSO can achieve global search, effectively control the error range, and make the results more reliable. In contrast, the GA-GRID method's scatter plot shows a more scattered and irregular distribution. While GA has strong global exploration capabilities, the convergence is less stable and results in greater randomness in errors. The SA-GRID method exhibits high density in local regions, which indicates that errors are concentrated within a limited range. However, the local density does not indicate strong overall optimization performance, suggesting that SA method is susceptible to local optima. SA lacks global consistency, leading to error concentration on a limited number of sensors, and its optimization performance is inferior to the global search capability and stability of PSO. In summary, the PSO-GRID method combines the advantages of global exploration and rapid convergence. The optimization results demonstrate stability and consistency, showing significantly better performance compared to the instability and dispersion of GA and the local concentration of SA.

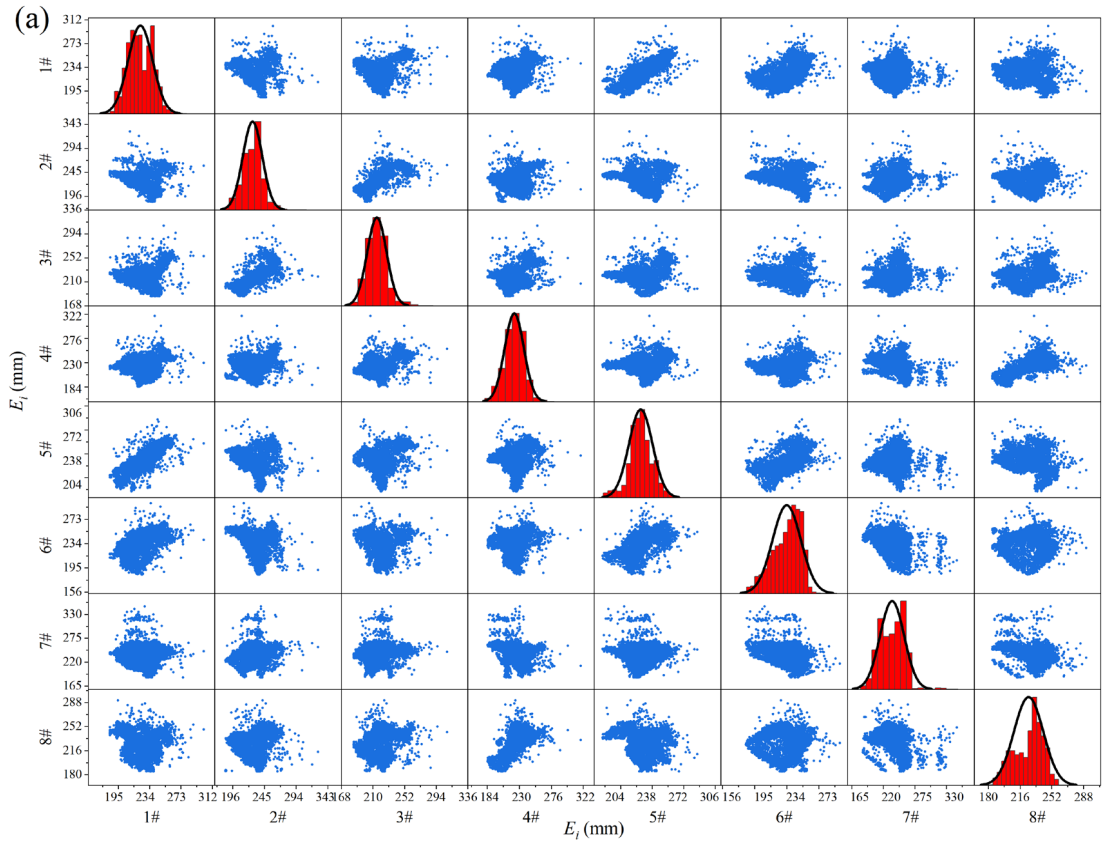

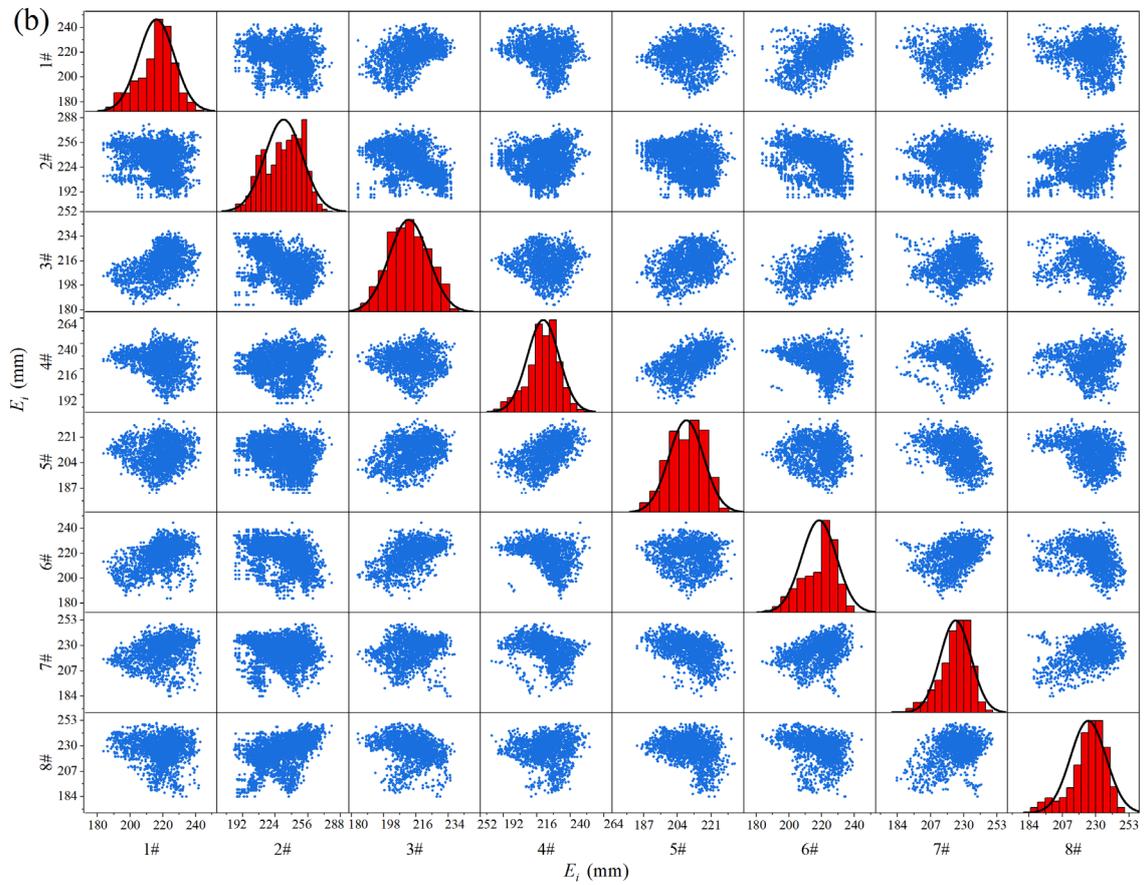

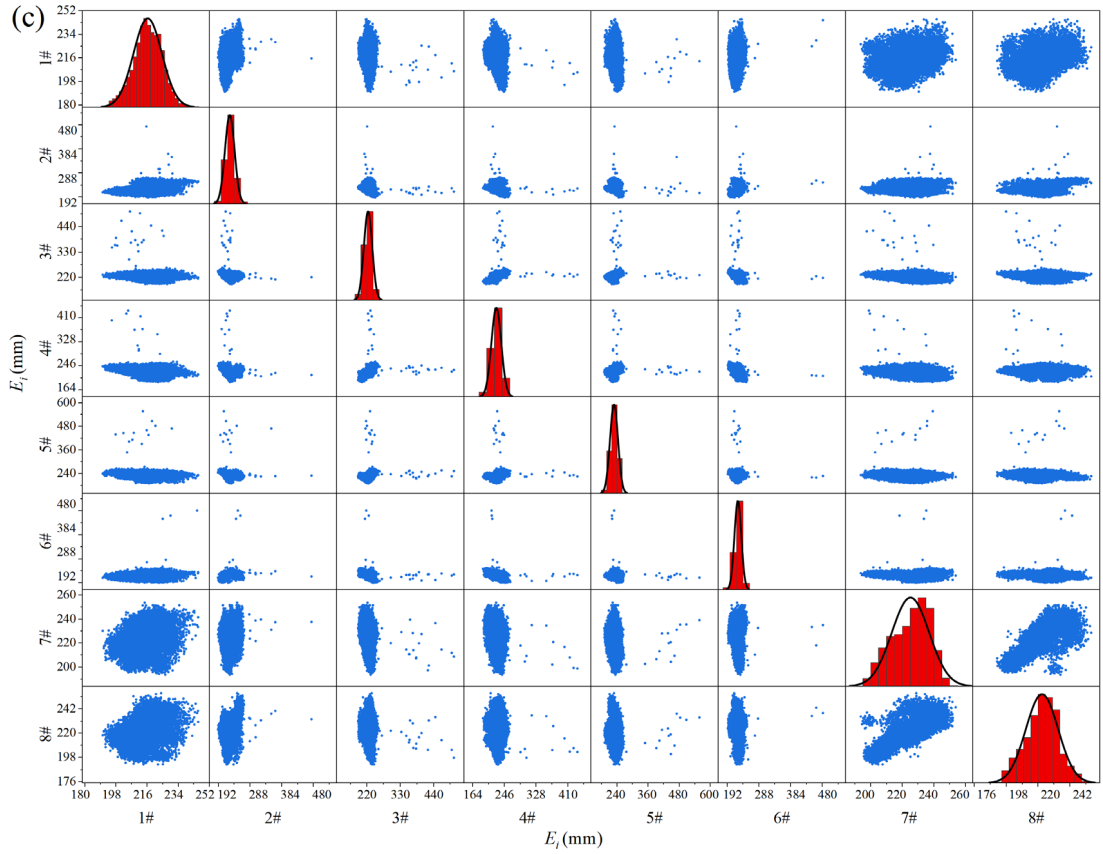

**Figure 6.** Scatter plots of the $E_i$ matrix of each sensor under different optimization algorithms. (a) PSO-GRID method. (b) GA-GRID method. (c) SA-GRID method.

The accuracy of AE event location is a critical indicator to evaluate sensor layout effectiveness, location errors are shown in Figure 7. The traditional layout has a relatively high average location error of 2.37 mm. After optimization using the PSO-GRID method, GA-GRID method, and SA-GRID method, the location errors decreased to 1.78 mm, 1.93 mm, and 1.92 mm. Compared with the traditional layout, the GA-GRID method, and the SA-GRID method, the PSO method improved location accuracy by 24.89%, 7.81%, and 7.39%, respectively. The results indicate that the PSO-GRID method achieves the best performance in reducing location errors, with the most significant improvement.

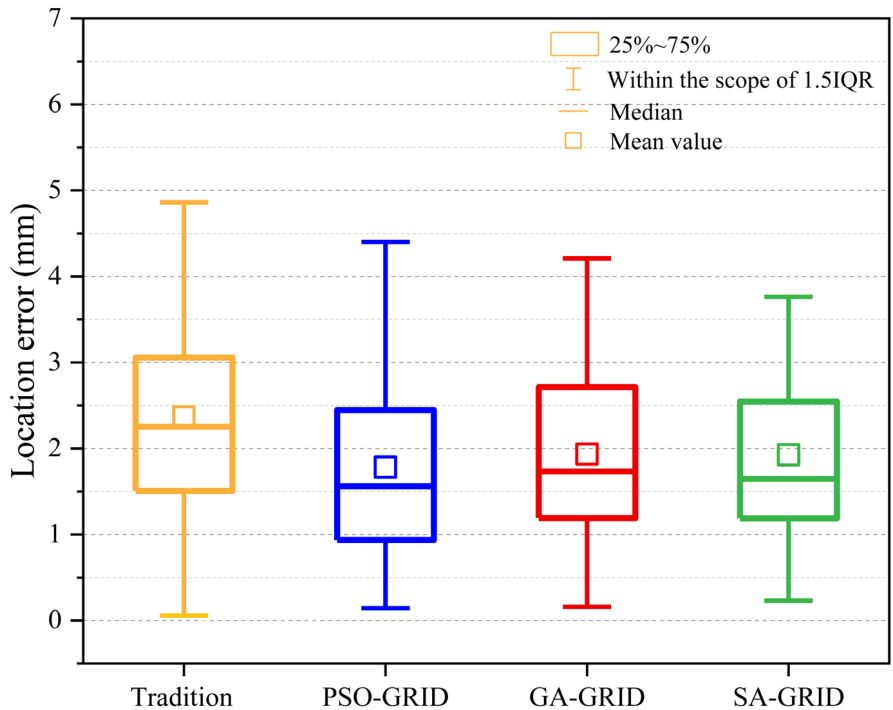

**Figure 7.** Box plot of location errors for traditional and different optimization algorithm layouts

**3.3 Noise resistance testing of the synthetic model**

In practical applications, sensors are inevitably influenced by noise, necessitating an evaluation of the proposed optimization method's adaptability to various noise environments. Gaussian noises with standard deviations of 0.5μs, 2μs, and 5μs were applied to the theoretical P-wave travel-time data. The location results are shown in Figure 8, where it is evident that the PSO-GRID method was the best performer, achieving location accuracy improvements of 24.89% (0.5 μs), 12.59% (2 μs), and 15.06% (5 μs) under these three noise conditions. This result confirms that PSO is better suited to handle high-noise data and provides a robust solution.

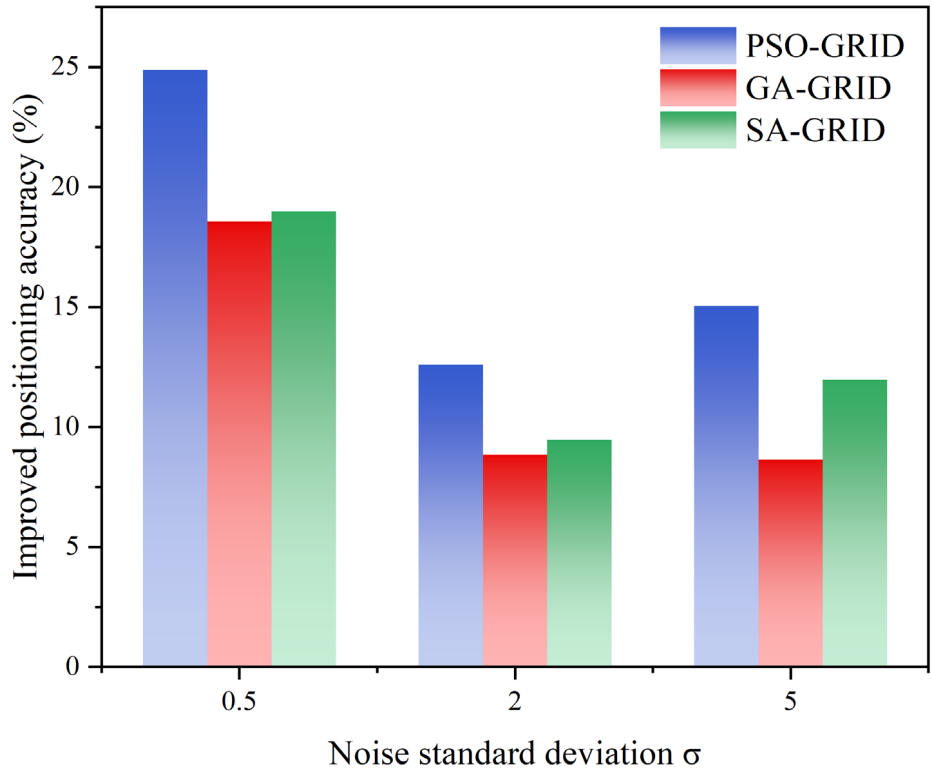

**Figure 8.** Changes in location accuracy under different noise parameters.

Additionally, Figure 9 generates a cumulative distribution function (CDF) plot based on location error. The PSO-GRID method optimization algorithm consistently delivers the best location accuracy improvement across all noise levels. Under low noise conditions (σ=0.5 μs), the cumulative probability curve for PSO is closer to the lower error range compared to the GA-GRID method and SA-GRID method. This indicates higher location accuracy and greater stability. As noise increases to medium and high levels (σ=2 μs andσ=5 μs), the PSO-GRID method still maintains strong noise resistance and shows a significant performance gap compared to the traditional layout. In contrast, the GA-GRID method and SA-GRID method gradually approach the performance of traditional layouts and become more sensitive to noise. Overall, the PSO-GRID method demonstrates the best location performance under various noise conditions. It is the preferred method for optimizing sensor layouts.

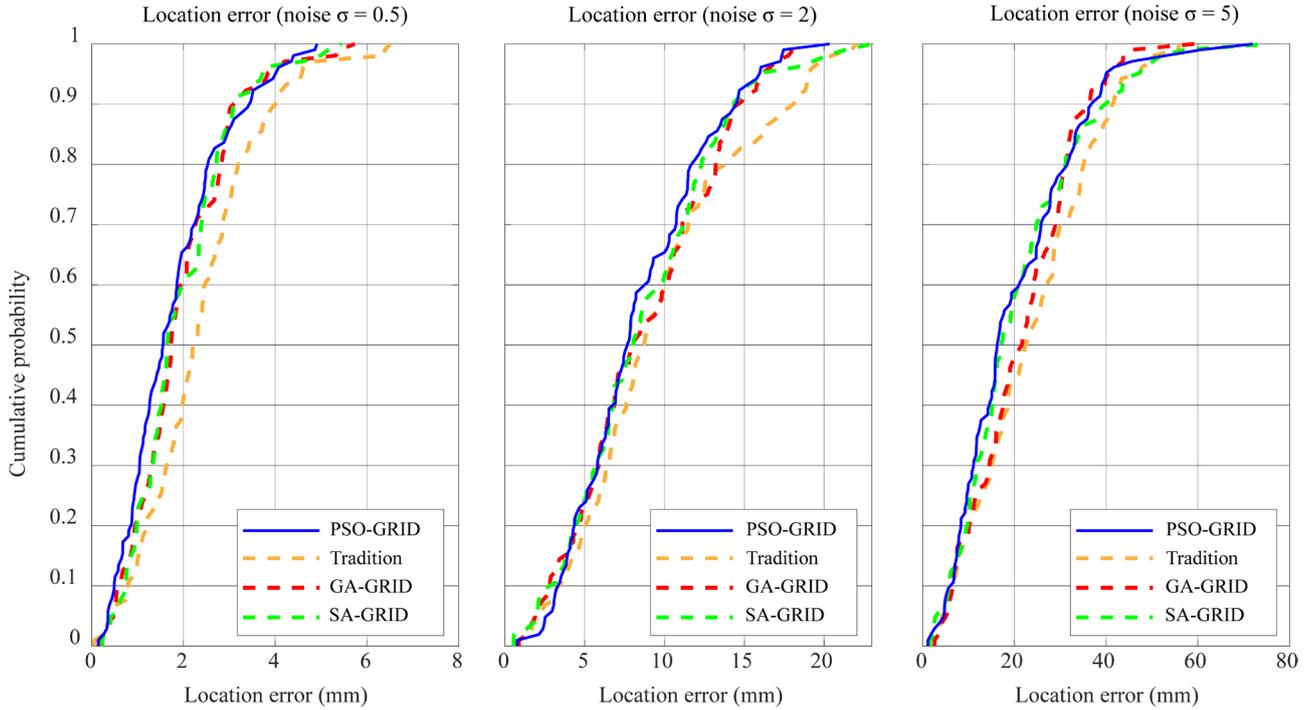

**Figure 9.** Cumulative distribution function of location error under different noise standard deviations.

## 4 Experimental test

### 4.1 Experimental setup

To further verify the effectiveness of the proposed method, this study conducted a laboratory AE PLB experiment on the traditional layout and the layouts optimized by the PSO-GRID method, GA-GRID method, and SA-GRID method. The DS2 AE equipment is made by the Softland Times company in Beijing (Figure 10). The dimensions of the granite sample and the P-wave velocity are consistent with the simulation test. The planar distribution of AE sources and sensors is shown in Figure 11. The sample surface was divided into 10 mm×10 mm grid units, with PLB performed at grid nodes to simulate AE sources. Due to operational constraints, some grid points near the sensors will not conduct PLB experiments.

In this experiment, 8 sensors were configured to capture signals emitted from the AE sources, all amplified by preamplifiers. The amplifier is fixed to 40 dB and the amplified signals were recorded by the DS5-8A device, with a sampling frequency set to 3 MHz. A total of 96 AE source points were used to collect signals for the optimized layouts of the traditional method, PSO-GRID method, GA-GRID method, and SA-GRID method. Each method utilized the same number of source points to ensure consistency in the comparison.

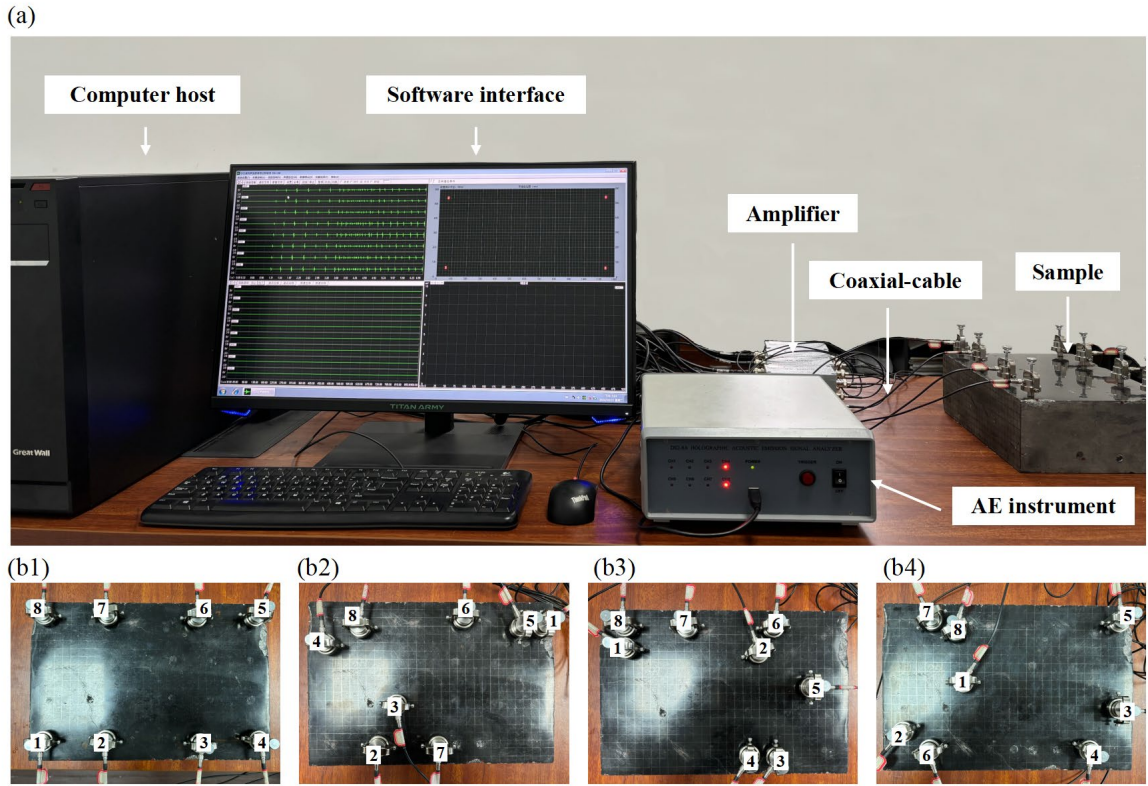

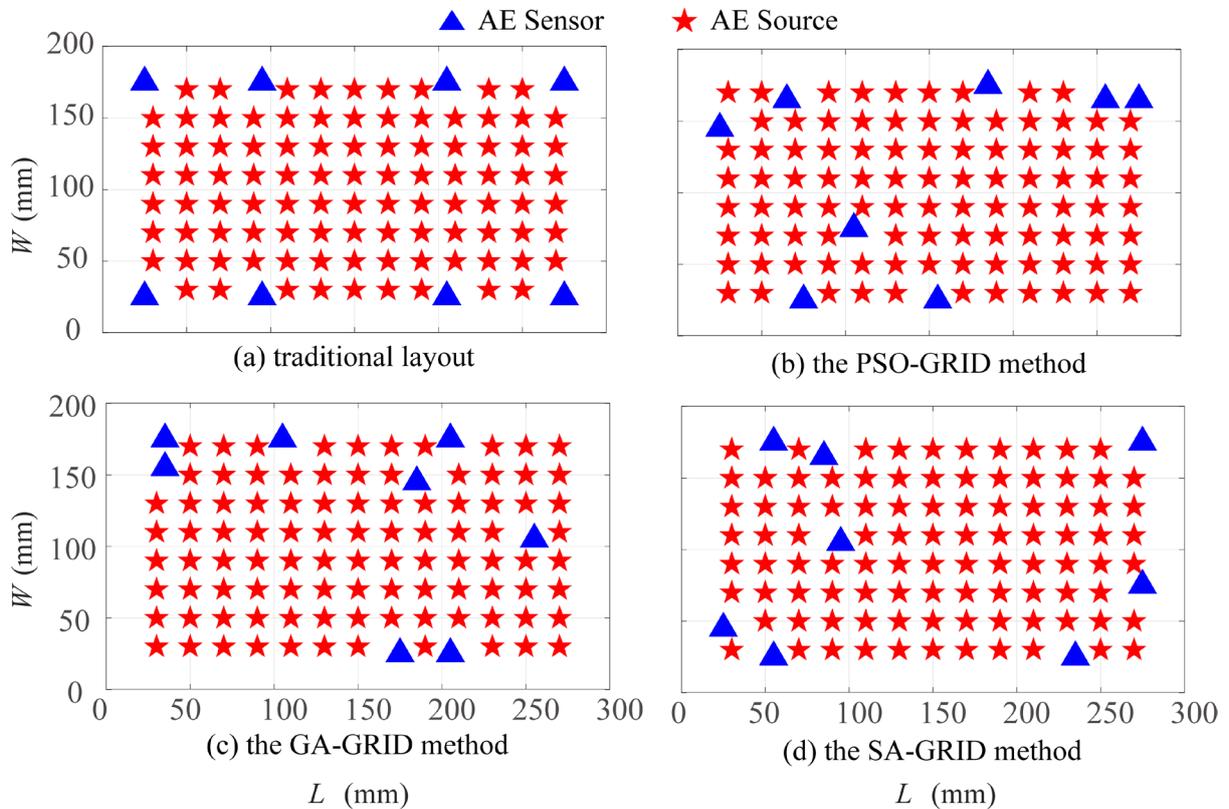

**Figure 10.** Detailed diagram of AE monitoring system and layouts. (a) AE equipment and sample. (b1) Traditional layout. (b2) PSO-GRID method layout. (b3) GA-GRID method layout. (b4) SA-GRID method layout.

**Figure 11.** Distribution of AE sources and sensor locations.

## 4.2 Experimental data processing

After the monitoring signals were received, they were analyzed using MATLAB software. The waveform of the AE signals received by the sensors is shown in Figure 12, with the P-wave first arrival data obtained using the AIC automatic picking method [18]. The use of the AIC method to pick P-wave arrivals is easily affected by trailing waves, which causes data instability. To address this issue, this study employs manual calibration and removes the origin time to ensure the accuracy of arrival time picking and subsequent location verification.

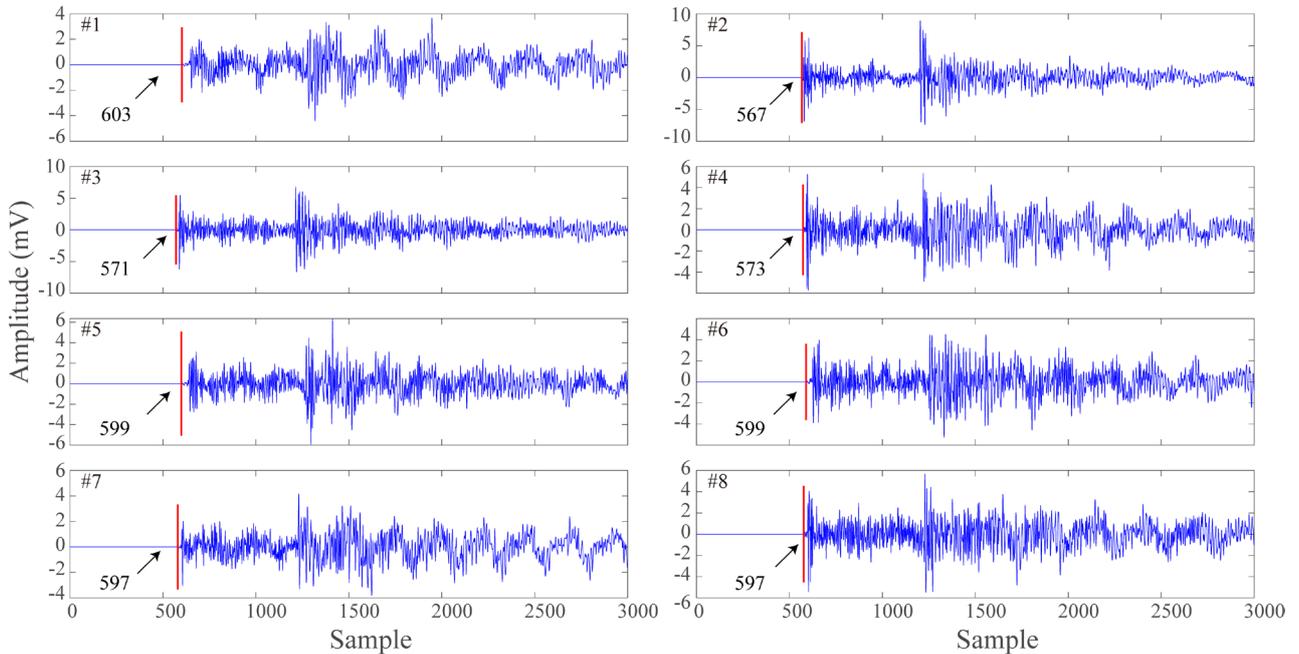

**Figure 12.** Waveform of PLB events and P-wave first arrival picking results. The red solid line indicates the picked P-wave arrival time. The number in the top left corner represents the sensor ID.

## 4.3 Location test

The layouts obtained from the synthetic tests in Section 3 were used in a location experiment, with the PLB points and AE location results shown in Figure 13b. The PSO-GRID method achieves the best performance in location accuracy. The error between the location result and actual location is generally small. The overall distribution of location errors is more uniform, and the location accuracy has significantly improved. In contrast, the GA-GRID method (Figure 13c) and SA-GRID method (Figure 13d) performed better than the traditional layout (Figure 13a). However, their location errors increase significantly in certain regions. This shows that GA and SA are more susceptible to traditional values.

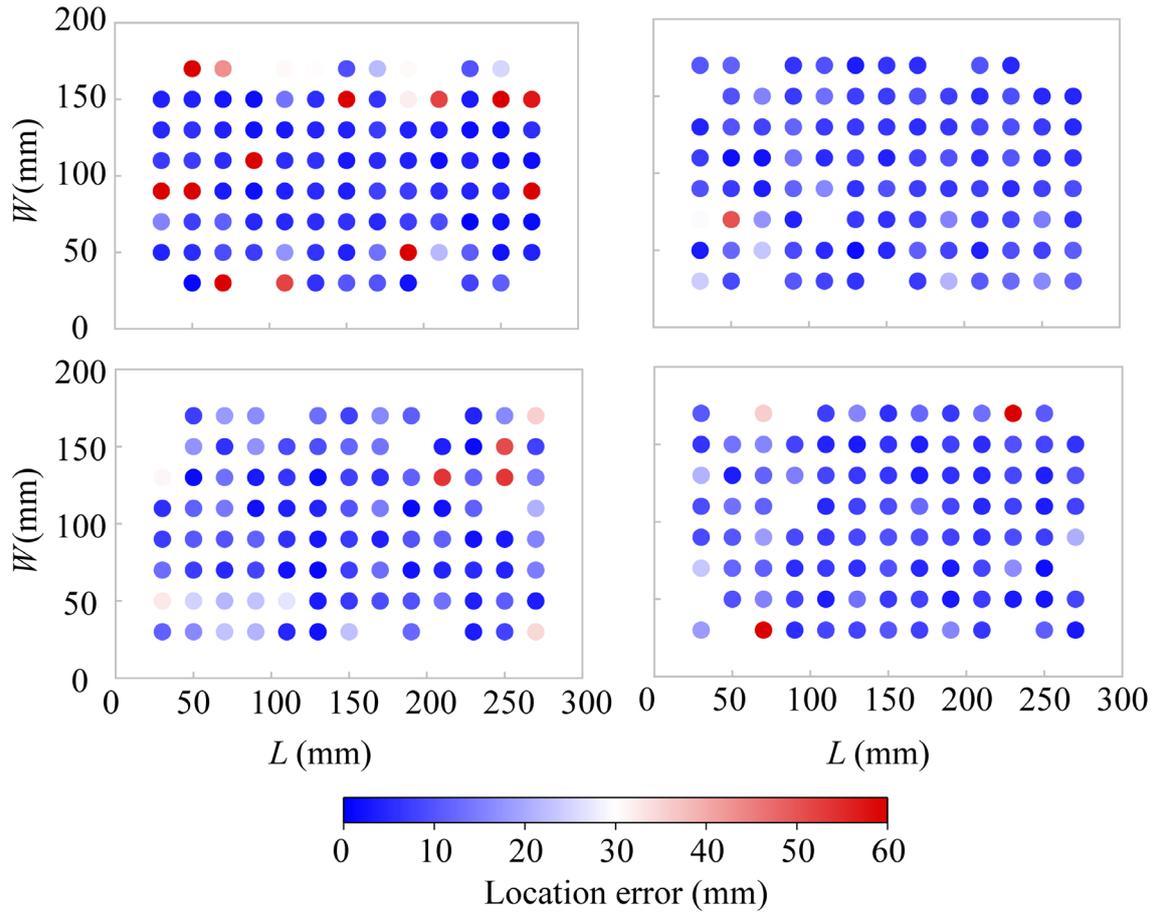

**Figure 13.** Location results of AE experiments under different layouts. (a) Traditional. (b) PSO-GRID. (c) GA-GRID. (d) SA-GRID.

The average location errors before and after optimization are shown in Figure 14. The average location error for the traditional layout is 22.94 mm, while the PSO-GRID method, GA-GRID method, and SA-GRID method optimized layouts achieved average location errors of 9.37 mm, 13.73 mm, and 11.78 mm. The proposed optimization method improved location accuracy by 59.15%, 31.76%, and 20.46% compared to the traditional layout, GA-GRID method, and SA-GRID method, respectively. This demonstrates the superior performance of the proposed method in enhancing location accuracy.

The PSO-GRID method demonstrates significant advantages in overall error, X-direction error, and Y-direction error. Compared with the traditional layout, the PSO-GRID method achieves a substantial reduction in error. This shows that it has the most notable optimization result. In comparison with the GA-GRID method and SA-GRID methods, the PSO-GRID method exhibits lower errors. In summary, the PSO-GRID method layout demonstrates significant advantages in accuracy and stability during the PLB experiment.

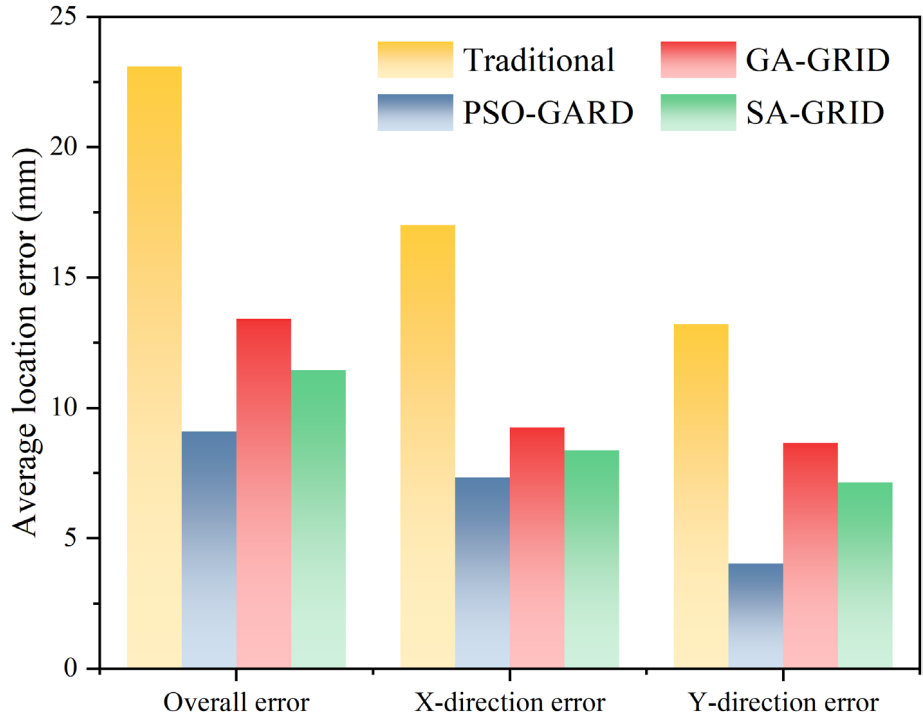

**Figure 14.** Box plot of overall average location error and average location error in X and Y directions under different sensor layouts.

## 5 Discussions

### 5.1 Influence of sensor layout

The determination of an optimal sensor layout is essential to ensure the smooth implementation of subsequent monitoring tasks in AE monitoring. Based on traditional empirical understanding, this study initially established a regular and partially enclosed sensor layout. However, through the optimization process, the sensor layout gradually evolved into a pattern where some sensors were distributed along the boundaries of the monitoring area, while others were located in the central region. Gomes [19] used a multi-objective optimization method to optimize the sensor layout. This approach focused on the maximization of the Fisher information matrix and the improvement of pattern reconstruction accuracy. The resulting layout was similar to the findings of this study.

This layout mainly results from a natural formation due to the optimization requirements of the objective function. To minimize the error between the predicted travel time and the actual travel time, the sensors must cover all potential seismic source points within the monitored area. Although arranging sensors solely along the boundaries can better envelop the monitored area, it will lead to increased errors in the distant central region. During the optimization process, the algorithm drives a small number of sensors to move from the boundaries toward the center to reduce errors in the central region. The combined effect of central and boundary sensors ensures a balance between global coverage and local accuracy, ultimately reducing the overall location error.

**5.2 Influence of optimization algorithms**

Sensor layout optimization is heavily influenced by the solution of the objective function. Selecting an appropriate algorithm to solve the objective function is key to the optimization process. PSO utilizes particle collaboration to achieve global exploration and prevent falling into local optima. Sadeq [20] applied PSO integrated with a triple-structure encoding method to optimize sensor placement in structural health monitoring. In sensor layout optimization, sensor locations are treated as continuous variables. GA's crossover and mutation operations may introduce redundancy, reducing search efficiency, whereas PSO does not have this issue and is more suitable for sensor layout optimization. Additionally, GA requires solutions to be encoded into discrete or binary forms, which can lead to decreased precision. In terms of computational efficiency, PSO is also more suitable for sensor layout optimization because its update equations are simple and the computational complexity is low. Additionally, the independent particle characteristics of the algorithm make it convenient for parallel processing. In contrast, the complex population operations of GA and the slow cooling process of SA make them difficult to achieve the efficiency required for engineering applications.

To further evaluate the performance of these three algorithms in AE location, we provide two types of analysis (Figure 15). The first is the fitness variation curves with iterations that show confidence intervals. The second is the total time distribution in bar charts for each algorithm. Figure 15(a) shows that all these three algorithms eventually converge to similar fitness levels. The GA algorithm exhibits larger fluctuations during convergence, which is due to the randomness of its genetic operations. In contrast, PSO and SA demonstrate greater stability, with SA exhibiting relatively low fitness values from the outset, attributed to its acceptance criteria. Figure 15(b) reveals that PSO has the shortest runtime, mostly concentrated between 0-1 s, while GA's runtime is more dispersed, mainly focused between 1-3 s. Although SA is more stable, its runtime is significantly longer, mainly concentrated between 6-8 s. The comprehensive analysis indicates that SA has advantages in stability but demonstrates low computational efficiency. In contrast, PSO achieves a good balance between efficiency and performance.

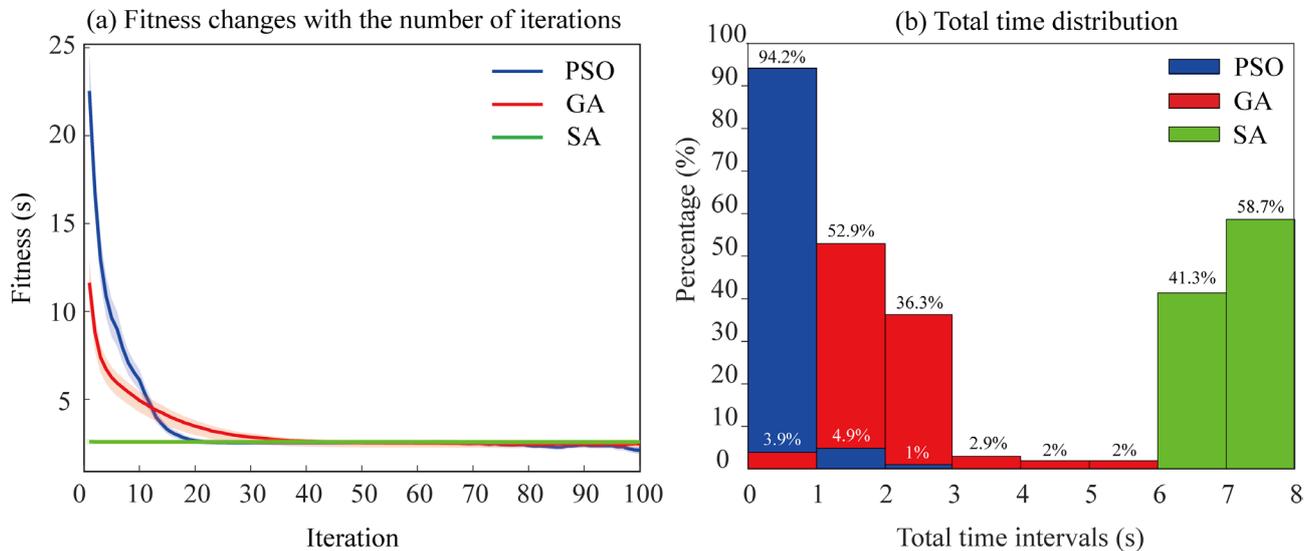

**Figure 15.** Comparison of fitness variation and time distribution.

## 6 Conclusions

To address the issues of traditional sensor optimization methods being difficult to understand and inefficient, a minimum travel-time-difference optimization method based on the PSO and grid search method is proposed. The method is compared with the GA-GRID method and SA-GRID method to verify the superiority of the proposed method. Finally, the feasibility and effectiveness of the method were verified through laboratory AE experiments. The main conclusions and findings are as follows:

(1) A sensor network layout optimization method based on the minimum travel-time difference is developed by combining the PSO method with the grid search method. The method establishes a location objective function based on the source travel-time theory and optimizes the function by PSO. It integrates grid search to systematically examine potential sensor locations, thereby determining the optimal sensor layout. This approach focuses on optimizing source location and is easy to understand.

(2) The optimization method proposed in this study demonstrated excellent location accuracy and strong noise resistance in synthetic tests. In the synthetic tests, the PSO-GRID method outperformed the traditional layout, GA-GRID method, and SA-GRID method, improving location accuracy by 24.89%, 7.81%, and 7.39%, respectively. Furthermore, under varying noise levels, the PSO-GRID method maintained high location accuracy, achieving improvements of 24.89%, 12.59%, and 15.06% across three noise conditions. These results validate the effectiveness and robustness of the proposed method in synthetic scenarios.

(3) The proposed optimization method performed well in laboratory AE experiments. In AE experiments, the PSO-GRID method improved location accuracy by 59.15%, 31.76%, and 20.46% compared to the traditional layout, GA-GRID method, and SA-GRID method, respectively. It also demonstrated high computational efficiency, minimal error fluctuations, and optimal performance under various noise conditions, requiring minimal parameter adjustment. These results confirm PSO-GRID as the most reliable and practical choice for location optimization.

**Funding:** This research was funded by State Key Laboratory of Strata Intelligent Control and Green Mining Co-founded by Shandong Province and the Ministry of Science and Technology, Shandong University of Science and Technology(No.SICGM2023301).